\documentclass[a4paper,fleqn,usenatbib]{mnras}

\usepackage{mathptmx}
\usepackage{txfonts}

\usepackage[T1]{fontenc}
\usepackage{ae,aecompl}

\usepackage{graphicx}	
\usepackage{amssymb}	

\title[Triangulum II for Dark Matter Detection]{The Potential of the Dwarf Galaxy Triangulum II for Dark Matter Indirect Detection}

\author[Genina \& Fairbairn]{Anna Genina \thanks{E-mail: anna.genina@kcl.ac.uk}
Malcolm Fairbairn\thanks{E-mail: malcolm.fairbairn@kcl.ac.uk}
\\
Physics, King's College London, Strand, London WC2R 2LS}

\date{Accepted XXX. Received YYY; in original form ZZZ}

\pubyear{2016}

\begin{document}
\label{firstpage}
\pagerange{\pageref{firstpage}--\pageref{lastpage}}
\maketitle


\begin{abstract}
  The recently discovered object Triangulum II appears to be an ultra faint dwarf spheroidal galaxy which may be one of the most dark matter dominated objects yet known.  In this work we try to estimate the potential of this object for studies of the indirect detection of self-annihilating dark matter by obtaining its astrophysical J-factor.  We perform a basic estimate of the velocity gradient to look for signs of the halo being tidally disrupted but show that the observed value is statistically compatible with zero velocity gradient.  We solve the spherical Jeans equation using Markov Chain Monte Carlo (MCMC) engine GreAT and the Jeans analysis part of the CLUMPY package.  We find the results point towards a very large J-factor, appearing to make Triangulum II one of the best targets in the search for dark matter. However we stress that the very small number of line of sight velocities currently available for this object make follow up studies essential.
\end{abstract}



\section{Introduction} \label{sec1}

N-body simulations of structure formation in $\Lambda$CDM universes suggest that galaxies such as our Milky Way should be surrounded by a number of dwarf galaxies with masses going down to below $10^6 M_\odot$ \citep{lcdm}.  Initially it was thought that the simulations were not a good mirror to nature, where many of these smaller haloes appeared to be missing \citep{missingsats}.  However, over the past decade, improved N-body simulations which include baryonic physics have shown that stellar feedback can empty such objects of baryons, making them more difficult to spot \citep{baryonicfeedback}.  At the same time, many dimmer dwarf galaxies have been discovered around our Galaxy by large surveys such as SDSS \citep{SDSS}.

Indeed over the past year or so, many new candidate dwarf galaxies have been found in the vicinity of the Milky Way by the Panoramic Survey Telescope and Rapid Response System (Pan-STARRS) \citep{kaiser} and the Dark Energy Survey (DES) \citep{DES}.  In total, more than 20 new Milky Way satellites have been discovered \citep{Leavens, bechtol, kimjerjen,drilca}.

The discovery of these objects has created a lot of excitetment in the astro-particle community because of their potential to help shed light on the nature of dark matter \citep{linden}.  In particular, in thermal relic dark matter theories such as WIMPs, where the abundance of dark matter is set by its self annihilation going out of equilibrium in the early Universe, we expect to be able to detect the products of dark matter annihilating with itself in galactic halos today \citep{lars}.  Dwarf Spheroidal galaxies, with their very high dark matter to baryon ratio, provide an excellent place to look since the signals of such annihilations would be less likely to be contaminated by more conventional backgrounds of baryonic origin (for example compact objects which can also emit gamma rays).  Searches for such signals in dwarf spheroidal galaxies with the Fermi gamma ray telescope are already maturing \citep{fermi} and the best targets for study will have to be chosen for the upcoming Cherenkov Telescope Array, which operates at higher energies \citep{CTA} and, unlike Fermi, requires a pointing strategy.

One of these objects, Triangulum II, has recently been found by the Pan-STARRS Survey \citet{Leavens} who, using only photometric data, were unable to ascertain whether the object was an ultra-faint dwarf galaxy or a globular cluster.  Follow up spectroscopic observations made by \citet{hotstellar} and \citet{kirby} disfavoured the Globular Cluster hypothesis and Triangulum II appears to be a dwarf galaxy located 36 kpc away from the galactic center.  In total, only 13 member stars have been studied in detail by one group \citep{hotstellar} and six stars by another \citep{kirby}, so any conclusions which are drawn at this stage about its dynamics will be preliminary. However, both groups observe an apparent low metallicity and a high velocity dispersion suggesting a large mass to light ratio for this object, both factors being consistent with the interpretation that this is a dwarf galaxy.  Triangulum II seems to exhibit complex kinematics, as the velocity dispersion appears to significantly increase from the central 10 pc outwards, which may suggest that the system is tidally disrupted. Signs of tidal disruption have previously been observed in Milky Way satellites \citep{bootes} \citep{segtid}. Despite this, the lack of visible ellipticity or non-Gaussianity in velocities appears to oppose the tidal disruption scenario \citep{kirby}.

In what follows, we will assume that these stars are reliable tracers of an underlying dark matter halo and calculate the J-factor - the relevant quantity which gives the integrated density squared of dark matter along the line of sight and over a solid angle $\Delta \Omega$, and consequently indicates the potential of a given dark matter halo to be a good target to search for the self-annihilation of dark matter.

First, in Section \ref{sec2}, we will remind the reader of the nature of the object, the distribution of stars and the velocity dispersions in different regions.

We will then search for a systematic variation in velocity across the object, which may be a signal of tidal disruption that would weaken the relevance of our J-factor calculation.

In Section \ref{sec3} we will outline the method we will be using for Jeans analysis of Triangulum II, necessary in further calculation of the J-factor, and test it on a simulated data set along with alternative methods to compare the performance of the software we are using to simpler techniques. In Section \ref{sec4} we will directly apply this method to Triangulum II to see how well or how badly we are able to constrain the parameters of its dark matter halo. Our results are then presented in Section \ref{jfac}.

As we were preparing the final version of this work, another paper appeared on the arXiv which set out to model the same galaxy \citep{scoop}.  The results of our analysis predict a J-factor which is very similar to their results.  We note that those authors also attempt to include the possible effects of triaxiality on the overall J-factor.  It is known that this can change the conclusions by a factor of a few \citep{bonni15}, however this would not prevent this object from being very interesting for particle physics phenomenology.

\section{Triangulum II} \label{sec2}

\begin{figure}

		\includegraphics[width=\columnwidth]{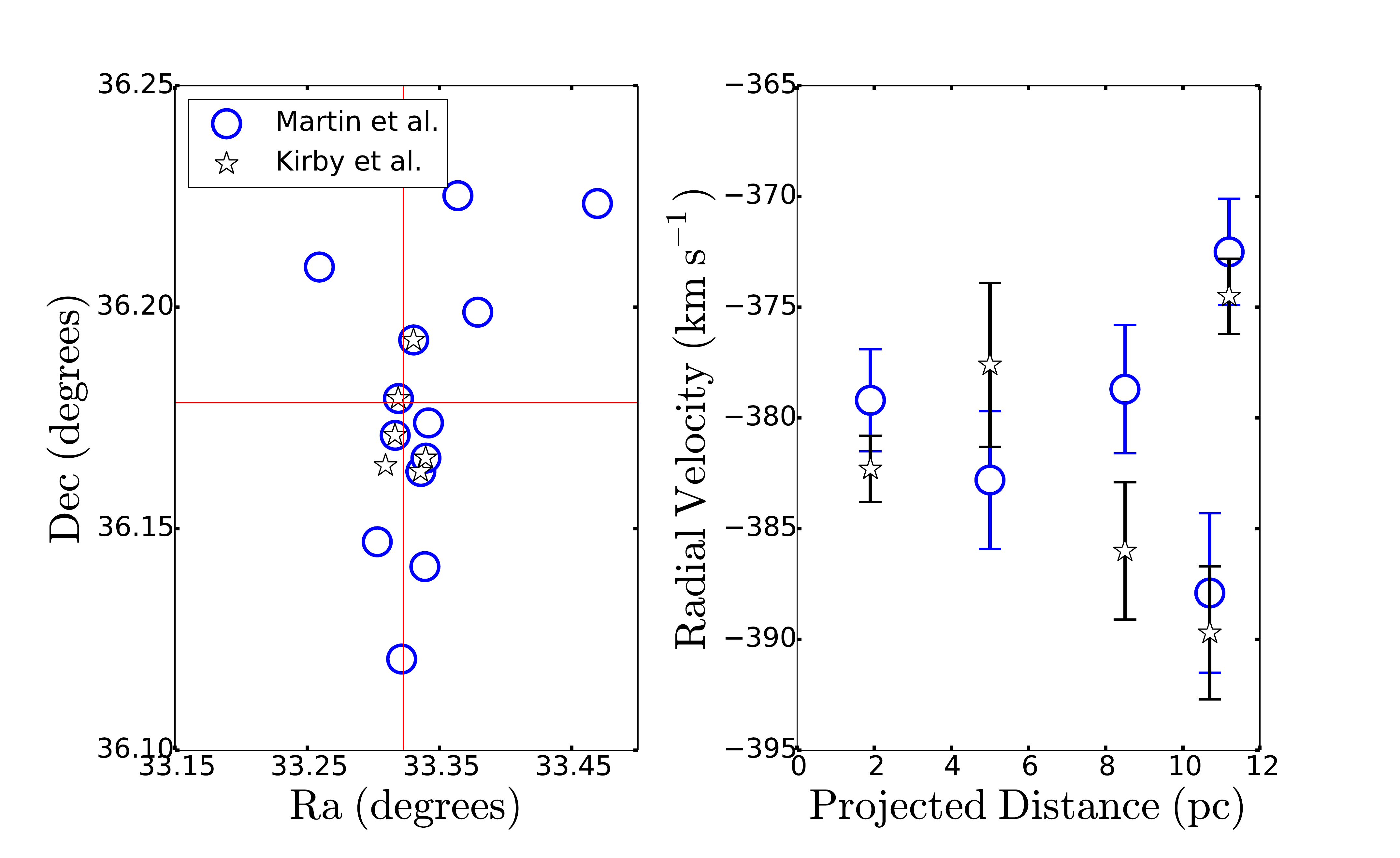}

		\caption{\it Left: Right Ascension and Declination of the stars measured by \citet{hotstellar} (blue circles) and \citet{kirby} (black stars). The red lines intersect at the location of Triangulum II centroid. Overlap in five of the stars from the two sets can be seen. Right: Velocity and velocity error measurements for the five overlapping stars by \citet{hotstellar} and \citet{kirby}. All but one of the measurements are consistent within error bars. }
                \label{starvels}

	\end{figure}

\begin{table}

		\begin{center}

                  \caption{Some properties of Tri II, $^{a}$ from \citet{Leavens}}
                  \label{tab:data}

		\begin{tabular}{c|c}
        \hline
			Heliocentric distance $^{a}$ & $30 \pm 2 $ kpc  \\

			$r_{1/2}$ $^{a}$ & $34^{+9}_{-8}$ pc \\
            
            $\langle v_r\rangle$ & $-384.3\pm{6.5}$ km s$^{-1}$ \\
            
			\hline
           
			Global Kinematics & \\
			
			$\sigma_{vr}$ & $8.6\pm{0.2}$ km s$^{-1} $ \\
           
            \hline
          
			Inner Kinematics & \\
		$\sigma_{vr}$ & $4.8\pm{0.8}$ km s$^{-1} $ \\
            
            \hline
           			Outer Kinematics & \\
			$\sigma_{vr}$ & $13.0\pm{4.3}$ km s$^{-1} $\\
                        \hline
	    \end{tabular}
		\end{center}
	\end{table}

\begin{table}
\centering
\caption{\it Tri II member stars. $R$ - projected distances, v$_{r}$ - heliocentric velocities, $\delta$v$_{r}$ - velocity errors. The fifth column indicates whether numbers are from \citet{kirby} (K) or \citet{hotstellar} (M), (K\&M) show weighted averages from the two sets.}
\label{tab:stars}
\begin{tabular}{ccccc}
\hline
  ID & R(pc) & v$_{r}$ $($km s$^{-1})$ & $\delta$v$_{r}$ $($km s$^{-1})$ & Data set \\       
\hline
1 & 1.9 &-381.4 & 1.3 & K\&M \\
2 & 5.0 &-380.7 & 2.4 & K\&M \\
3 &8.5 &-382.1 & 2.1 & K\&M \\
4 & 10.2 &-384.9 &3.2 & K \\      
5 & 10.3 &-383.1 & 4.9  & M     \\
6 & 10.7 &-389.0 & 2.3 & K\&M \\
7 & 11.2 &  -373.8 & 1.4 & K\&M \\
8 & 19.4 &-387.0 & 3.8 & M       \\
9 & 21.2 &-401.4 & 6.6  & M     \\
10  & 30.3 &-362.8 & 5.6  & M    \\
11  & 31.4 &-397.1  &7.8  & M     \\
12 & 32.7  & -404.7 & 5.1   &M    \\
13 & 36.8 & -387.1 & 7.7  &M     \\
14  & 80.4 & -375.8 & 3.1 & M    \\  
\hline

\end{tabular}
\end{table}

As mentioned in the introduction, Triangulum II appears to be a metal poor dwarf galaxy with a large mass to light ratio.  Only 13 member stars have been studied in detail by one group \citep{hotstellar} and six stars by a second group \citep{kirby}.  By mapping out the positions of the stars, we were able to determine that five of the stars overlap between the two sets and there are only 14 stars in total available for analysis. The overlap can be seen on the left hand side of Figure \ref{starvels}. For these five overlapping stars the right ascension and declination measurements from both sets of data were identical to 0.001 degrees, which, together with the overall pattern of the stars on the sky, suggests that these are indeed the same stars. The right hand side of Figure \ref{starvels} shows the measured velocities and corresponding errors for the identical stars in the two sets of data. All measurements but one (the star at 8.5 pc from the centroid) agree with each other within error bars. The disagreement in that particular star could indicate a calibration error in either instrument, however for the remainder of this work we obtain the weighted means for this star and the other overlapping stars from both sets of data. The positions in parsecs of each star were determined from right ascension and declination measurements of \citet{hotstellar} and \citet{kirby}, using the small angle approximation. Table \ref{tab:data} shows how the velocity dispersion varies as we look at stars closer and further from the centre of the galaxy.
	
We use an exponential profile to model the surface brightness of Triangulum II.
\begin{equation}
  \Sigma(R) = \Sigma_0 \exp (-\frac{R}{r_c})
  \label{explight}
\end{equation}
where $r_{c}$ is the scale radius \citet{los}. We used the probability density function for the half-light radius presented in Figure 3, left, of \citet{Leavens} to obtain a value for the scale radius of $r_c=r_{1/2}/1.68=21_{-5}^{+6}$ pc, with $r_{1/2}$ - the half-light radius and $1.68$ being a typical conversion factor between the half-light radius of a galaxy and the characteristic radius of an exponential profile. The normalisation $\Sigma_0$ is obtained in a similar fashion from the histogram of the magnitude in Figure 3, right, of \citet{Leavens}, which we approximate as a Gaussian probability density $\left(\mu,\sigma\right)$ = (-1.8, 0.5). We convert the measured magnitude in the visual band $M_V$ = 1.8$\pm$0.5 \citep{Leavens} to luminosity via the relation \citep{schneider}: 
\begin{equation}
\frac{L_V}{L_{\odot,V}} = 10^{0.4\left(M_{\odot, V} - M_V\right)}
\end{equation}

where L$_V$ is the luminosity and M$_V$ - magnitude in the V-band. We use M$_{\odot,V}$ = 4.83 \citep{binneymerrifield} and obtain $\log(L_V/L_{\odot,V})$ = 2.65$\pm$0.20, identical to that found by \citet{kirby}. Then, for an exponential profile, we compute the normalisation as: \citep{los} 
\begin{equation}
\Sigma_{0} = \frac{L_{tot}}{2\pi r_c^{2}}
\end{equation}
resulting in a value of $\Sigma_0=1.62_{-0.08}^{+0.15}\times 10^5$ $L_\odot$ kpc$^{-2}$, assuming a distance of 30$\pm$2 kpc. The central values are the values that we use in the analysis.

Table \ref{tab:stars} lists the stars that we use for our analysis, their distance from the centre of the object and the line of sight velocities obtained from spectroscopy with their respective errors.  It can be seen that Triangulum II is moving very quickly towards the Milky Way, which may also suggest that one should check to see if it is in the process of being tidally disrupted.

It has long been known that streams of stars, which result from star clusters being tidally disrupted as they move through the potential of the Milky Way, can appear to look like dwarf spheroidal galaxies if they are oriented parallel to the line of sight \citep{falsealarm}.  With so few stars it is quite difficult to test as to whether this dwarf is a self-gravitating, static (or very nearly static) solution of the Jeans equation or whether it is in the process of being tidally disrupted.  One simple test, which would indicate whether such a process might be under way, is to look at the velocity gradient across the halo to see if one side is moving with a sigficantly different velocity to the other.  In particular, if the velocity gradient times the size of the halo is larger than the velocity dispersion of the halo, this could be a strong indication that the halo is breaking up \citep{pryor}.

\begin{figure}

		\includegraphics[width=\columnwidth]{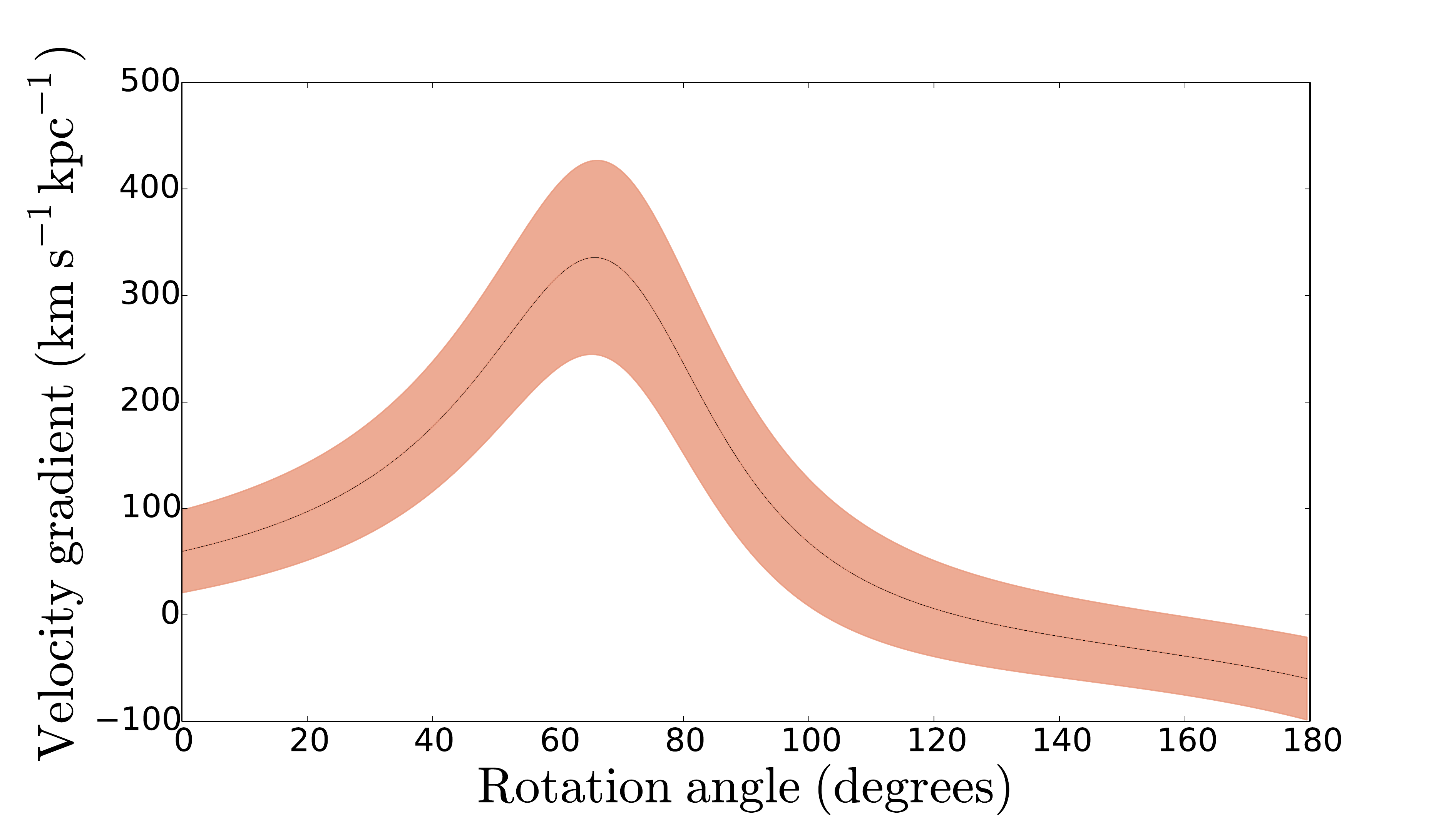}

		\caption{\it Velocity gradient across the halo for different orientations.  The width of the band reprsents the errors from linear regression fitting while the central line is the best fit value of the gradient.  The maximum value is $335.7 \pm 91.2$ km s$^{-1}$ kpc$^{-1}$.}

		\label{velgrad}
	\end{figure}

  In order to investigate this, we rotate the stars around the line of sight through the centre of the stellar distribution and for each orientation we note the line of sight velocities from left to right and carry out a linear least squares fit to data, searching for the orientation which maximises the velocity gradient across the halo.  Figure \ref{velgrad} shows that indeed there does appear to be a larger gradient for one particular orientation.

\begin{figure}

		\includegraphics[width=\columnwidth]{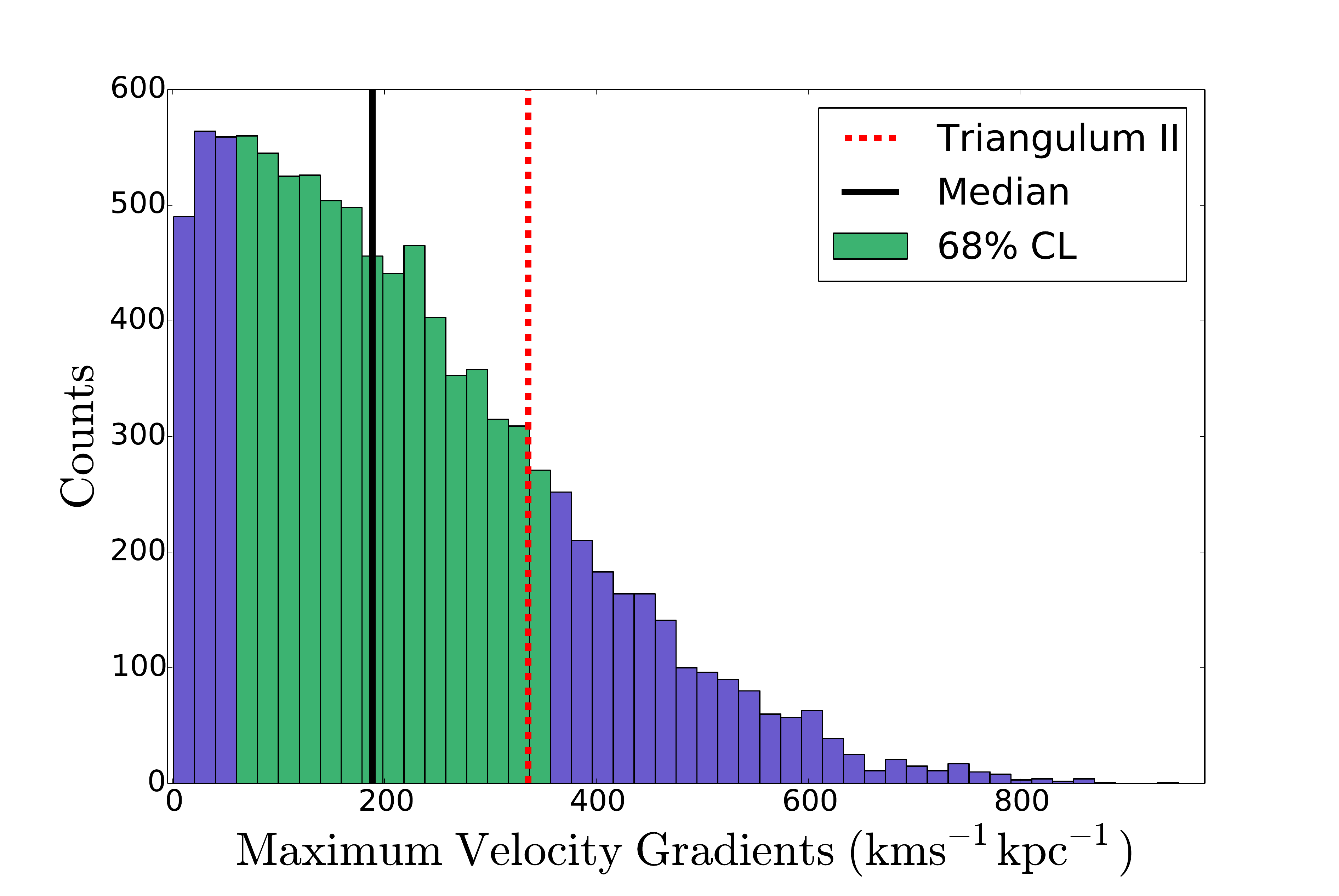}

		\caption{\it Histogram of Velocity gradients for 10,000 realisations of the same distribution of stars with random line of sight velocities drawn from the observed velocity distribution of the halo. The found value of $335.7\pm 91.2$km s$^{-1}$ kpc$^{-1}$  is therefore not statistically significantly different from no gradient at all and lies within 68\% CL of the median velocity gradient ($188.6$km s$^{-1}$ kpc$^{-1}$) generated by velocity dispersion alone.}
                \label{hist}

	\end{figure}
Naively, the magnitude of the velocity gradient at that orientation ($335.7 \pm 91.2$ km s$^{-1}$ kpc$^{-1}$) should also be a cause for concern (although the error on this is very large).  The largest distance of the 14 stars from the centre of the halo is 80.4 pc and for a diameter of 160.8 pc one would expect a velocity difference across the halo due to such a gradient of around 54 km s$^{-1}$ which is much larger than the velocity dispersion of 13.0 km s$^{-1}$.

However, for such a small number of stars, we need to be very careful as we would expect large fluctuations in the average line of sight velocity in a particular region of ths star field.  To find out the statistical significance of this velocity gradient we therefore perform the following test:-  We take the x,y position of the stars and the errors on their velocities, but we generate fake random values for the line of sight velocities, based upon the velocity dispersion measured for the inner group and the outer groups of stars as shown in Table \ref{tab:data}.  For each realisation, we then perform the same test we did with the real data and rotate it, obtaining the orientation with the maximum velocity gradient.  We then add the maximum velocity gradient for each realisation to a histogram, which is shown in Figure \ref{hist}.

The results of this analysis show that the value of  $335.7\pm 91.2$ km s$^{-1}$ kpc$^{-1}$ is within the 68\% confidence interval - so is a very typical value, completely consistent with the random velocity gradient one would obtain from a distribution with no true velocity gradient at all, when sampled by so few stars.

The fact that such a large velocity gradient, much larger than the velocity dispersion of the halo, is still statistically insignificant really demonstrates the paucity of data available for this object.  While we have no evidence that tidal disruption is taking place as such, we also have distinct evidence that more data is required before we can say anything with too much confidence.

Nevertheless, we will proceed to calculate the J-factor for this object.

\section{Cross-check of CLUMPY Against Other Methods} \label{sec3}

In this section, we will describe the tools and procedures that we use to fit the gravitational potential in which the stars are moving and the steps we took to test these tools by cross-checking them with simpler methods on mock data.

For our main result, we use the MCMC analysis toolkit GreAT, included in the CLUMPY package, to solve the spherical Jeans equation \citep{GreAT,CLUMPY}.  The likelihood function in CLUMPY for performing the Jeans analysis is set to be a product of likelihoods for individual stars.  We assume a spherically symmetric system of collisionless particles, acted upon by the gravitational potential of dark matter:
\begin{equation}
\frac{1}{\nu} \frac{d}{dr} (\nu \sigma^{2}_{r}) +  2 \frac{\beta_{ani}(r) \sigma^{2}_{r}}{r} = -\frac{GM(r)}{r^2}
\end{equation}
where $\nu(r)$ is the 3D light profile, $\sigma_r$ is the radial component of velocity anisotropy and $\beta_{ani} = 1 - (\frac{\sigma_{\theta}}{\sigma_{r}})^2$ with $\sigma_{\theta}$ - the tangential velocity dispersion.

As a verification of our approach and of the CLUMPY code,  we compare the resuts of the code to those obtained using the projected virial theorem.  In \citet{Kent}, it is shown that the classical virial theorem may be projected in 2D
  \begin{equation}
    \int_{0}^{\infty} \Sigma\langle v^{2}_{z}\rangle RdR = \frac{2}{3} \int_{0}^{\infty} \nu \frac{d \Phi}{dr} r^3 dr
	\end{equation}
where $\langle v^{2}_{z}\rangle$ is the second line-of-sight velocity moment, $\Sigma$ is surface brightness and $\nu$ is the 3D light profile.  For valid solutions, the weighted integral over the velocity dispersion on the left hand side will be equal to the weighted integral over the derivative of the gravitational potential on the right.  We followed the procedure outlined in \citet{Richardson} to implement the projected virial theorem.

Strictly speaking, both the solution of the Jeans equation via CLUMPY and the projected virial theorem should yield a similar result, but since they are coded seperately they are a useful way to check for problems with the code or our implementation of that code.

\begin{figure}

		\includegraphics[width=\columnwidth]{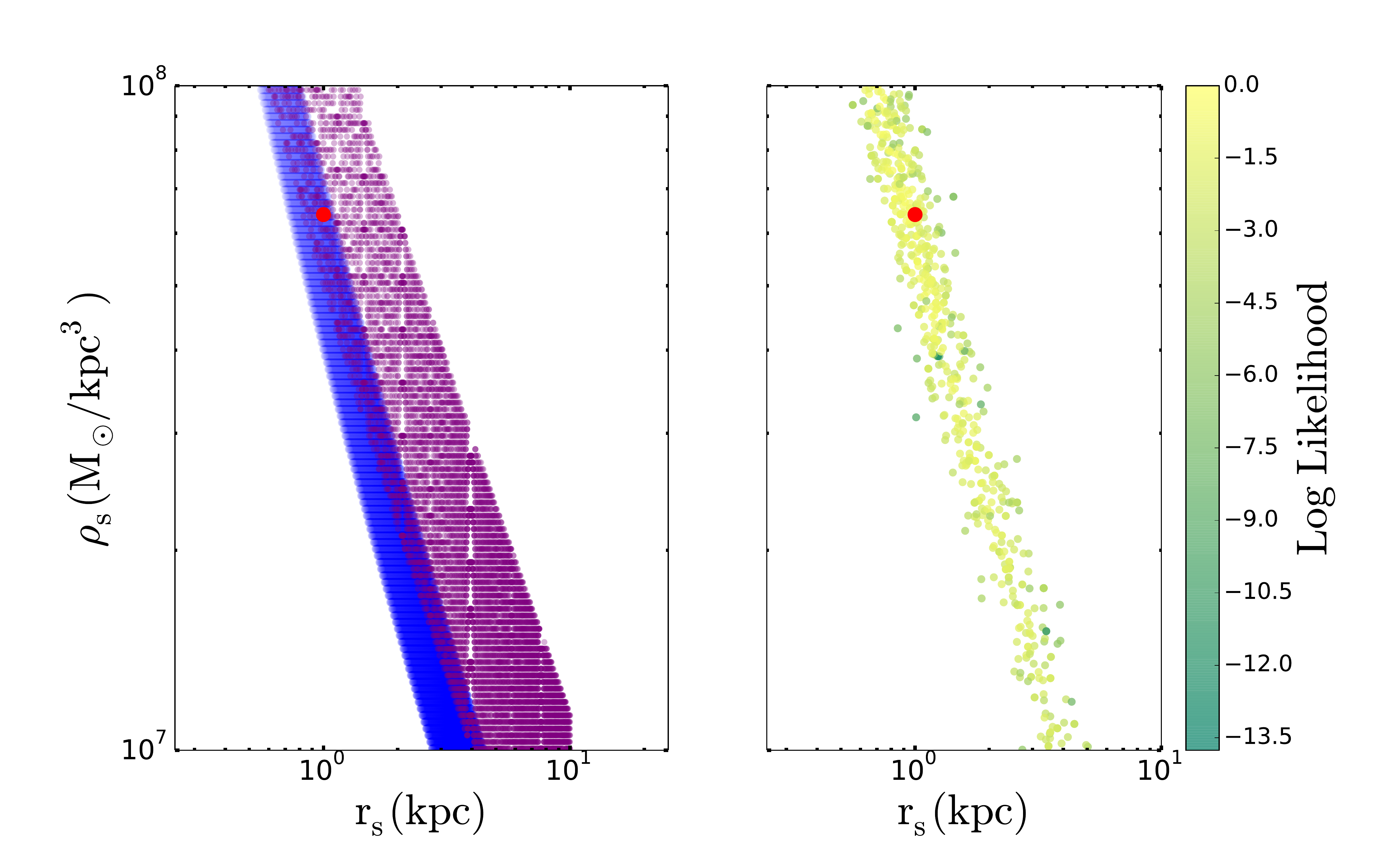}

		\caption{\it Constraints on $\rho_s$ and $r_s$ of the simulated galaxy from the Gaia challenge data suite.  On the left the best fit parameters from the projected virial theorem are shown as a blue dots and the parameters from the mass at half light radius method are the purple dots. The "real" point with known parameters is shown as a red dot.  On the right we estimate the same parameters using CLUMPY.  The Log-likelihood is normalised with respect to the point with best likelihood. }

		\label{fake}
	
	\end{figure}

We also implemented the method of \citet{Wolf}, who approximate the mass contained within half-light radius of dispersion-supported galaxies,  derived from the Jeans equation. It is suggested that at the radius approximately equal to the 3D deprojected half-light radius the obtained mass is insensitive to dispersion anisotropy $\beta_{ani} = 1 - (\frac{\sigma_t}{\sigma_r})^2$, which is extremely advantageous since only one component of the velocity dispersion can be measured and assumptions normally have to be made about $\beta_{ani}$.  The following relation is derived :
    \begin{equation}
    M_{\frac{1}{2}} = 3G^{-1}\langle\sigma_{p}^{2}\rangle r_{1/2} \simeq 4G^{-1}\langle\sigma_{p}^{2}\rangle R_e
    \end{equation}
    where $G$ is the gravitational constant and $R_e$ is the 2D projected half-light radius. The relation holds if the line-of-sight velocity dispersion $\sigma_{p}$ is approximately flat at the half light radius. \\
    The luminosity weighted line-of-sight velocity dispersion is defined as \citep{los}:
    \begin{equation}
    \langle\sigma^2_p \rangle = \frac{\int_0^\infty \sigma^2_p (R) \Sigma(R) R dR}{\int_0^\infty \Sigma(R) R dR}
	\end{equation}
    All three methods were tested on fake data from the Gaia Challenge suite of simulations \citep{gaiawiki, gaia} to ensure that not only were the three approaches consistent with each other, but that they were also consistent with the known underlying gravitational potential - in this case that potential resulting from an NFW density profile
    \begin{equation}
      \rho(r)=\frac{\rho_s}{\frac{r}{r_s}\left(1+\frac{r}{r_s}\right)^2}
     \end{equation} 
    with the scale density $\rho_s = 6.4 \times 10^{7}$ $M_{\odot}$ kpc$^{-3}$  and the scale radius $r_s = 1.0$ kpc - and a Constant stellar anisotropy profile.  We split 100 stars of the mock data into 10 bins of ($\sim$ 8-10) stars. The velocity of each star was assigned a random error drawn from the error distribution of \citet{hotstellar} data. For each bin, we find the estimates
    of the velocity dispersion and the corresponding errors using maximum likelihood statistics, the procedure for which is outlined in \citet{walkererrors}. We then use the Projected Virial Theorem and the method of Wolf et al. to find values of $\rho_s$ and $r_s$ and compare to the actual value, this is plotted on the left hand side of Figure \ref{fake} which shows the two methods are consistent with each other within the error bars.

    We obtained velocity and position data for a 100 simulated stars, which we input into CLUMPY. We set our priors to [5 , 13] for log[$\rho_s$] and [-1.4 : 1] for log[$r_s$] (starting at the half-light radius of 250 pc) and run the MCMC. The result of this procedure can be seen on the right hand side of Figure \ref{fake} and shows that all three methods are consistent with each other and the true point.
In this way we were able to verify that the CLUMPY code yields results that are consistent with other analyses and is able to reproduce the correct underlying parameters when applied to mock data sets.

\section{Jeans Analysis for Triangulum II} \label{sec4}

In this section we describe the assumptions that we made while using CLUMPY to fit the underlying density profile of Triangulum II.

Throughout the Jeans analysis on Triangulum II we assume a generalised Navarro-Frenk-White profile for the distribution of dark matter, (sometimes known as a Zhao \citep{zhao} or $\alpha\beta\gamma$ profile)
\begin{equation}
\rho(r)=\frac{\rho_s}{\left(\frac{r}{r_s}\right)^\gamma\left[1+\left(\frac{r}{r_s}\right)^\alpha\right]^{\frac{\beta-\gamma}{\alpha}}}
\end{equation}
where $\rho_s$ is the characteristic density parameter, $r_s$ is the scale radius, $\gamma$ gives the logarithmic density gradient at small radii and $\beta$ at large radii. The third power $\alpha$ controls the rate at which the density profile interpolates between its inner and outer values.

As a fast moving satellite of the Milky Way on a quite radial velocity path, Triangulum II is likely to have undergone eccentric orbits. \citet{tidal}, using N-body simulations, show that a satellite orbiting the Milky Way would lose a fraction of its mass after every pericentric passage, and the smaller is the pericenter of the orbit the stronger are the tidal effects. It is shown that the outer mass profile of a tidally stripped halo is well approximated by an outer slope of $\propto r^{-4}$. The orbit of Triangulum II is currently unknown. Although the lack of visible ellipticity or non-Gaussianity in velocities as well as the velocity gradient test carried out in Section \ref{sec2} all appear to oppose the tidal disruption scenario \citep{kirby}, the possibility of tidal stripping of dark matter in the outer parts of the galaxy is not excluded. Whilst the velocity dispersion profile out to the furthest star may be representative of Triangulum II kinematics, the assumption that the dark matter halo in the regions outside the furthest known member star would match that of an isolated halo, far from a large galaxy like the Milky Way, is largely unmotivated.

We have therefore considered halos of the form $\beta=4$ while the other parameters $\alpha$ and $\gamma$ were let free, as well as allowing all three parameters to vary.  We found the choice of $\beta=4$ did not make a huge difference on J-factor estimates since the integral of the J-factor is more sensitive to the profile at lower radii.  We therefore allowed $\alpha$, $\beta$ and $\gamma$ to vary in the final fit.

For the stellar density $\nu(r)$ we adopt the functional form which fits with the projected density profile fitted to the data in Equation \ref{explight}.  We adopt a Constant stellar anisotropy profile $\beta_{ani}(r)=\beta_{ani0}$, where we keep the constant $\beta_{ani0}$ a free parameter.

The unbinned velocity data in Table \ref{tab:stars} has been used as input and the priors for MCMC are shown in Table \ref{tab:priors}, as suggested for dwarf spheroidals in \citet{Input}. Note that the minimum value of $r_s$ has been set to be above the half-light radius of the system, thus implying that the halo must be at least as large as the volume enclosed within that half-light radius. It is discussed in \citet{Input} that too low values of $r_s$ lead to high density at the center of the halo, resulting in J-factors that are too high. Calculation of J-factors is discussed in the next section. \\
We performed the MCMC analysis using 10 chains with 10000 points per chain.
\begin{table}

		\caption{Triangulum II MCMC priors for the Dark Matter profile and Constant stellar anisotropy profile}
		\label{tab:priors}
\begin{center}
		\begin{tabular}{c|c}
                  \hline
		  $\log_{10} \rho_{s}$ & [5, 13] \\

			$\log_{10} r_{s}$ & [-1.4, 1] \\

			$\alpha$ & [0.12, 1]\\

			$\beta$ & [3, 7]\\

            $\gamma$ & [0, 1.5]\\
                  
                  \hline
           
                  $\beta_{ani0}$ & [-9, 1] \\
                  \hline
		\end{tabular}
\end{center}	
	\end{table}

\begin{figure}

		\includegraphics[trim=1.5cm 1.5cm 2.5cm 0.5cm, width=\columnwidth]{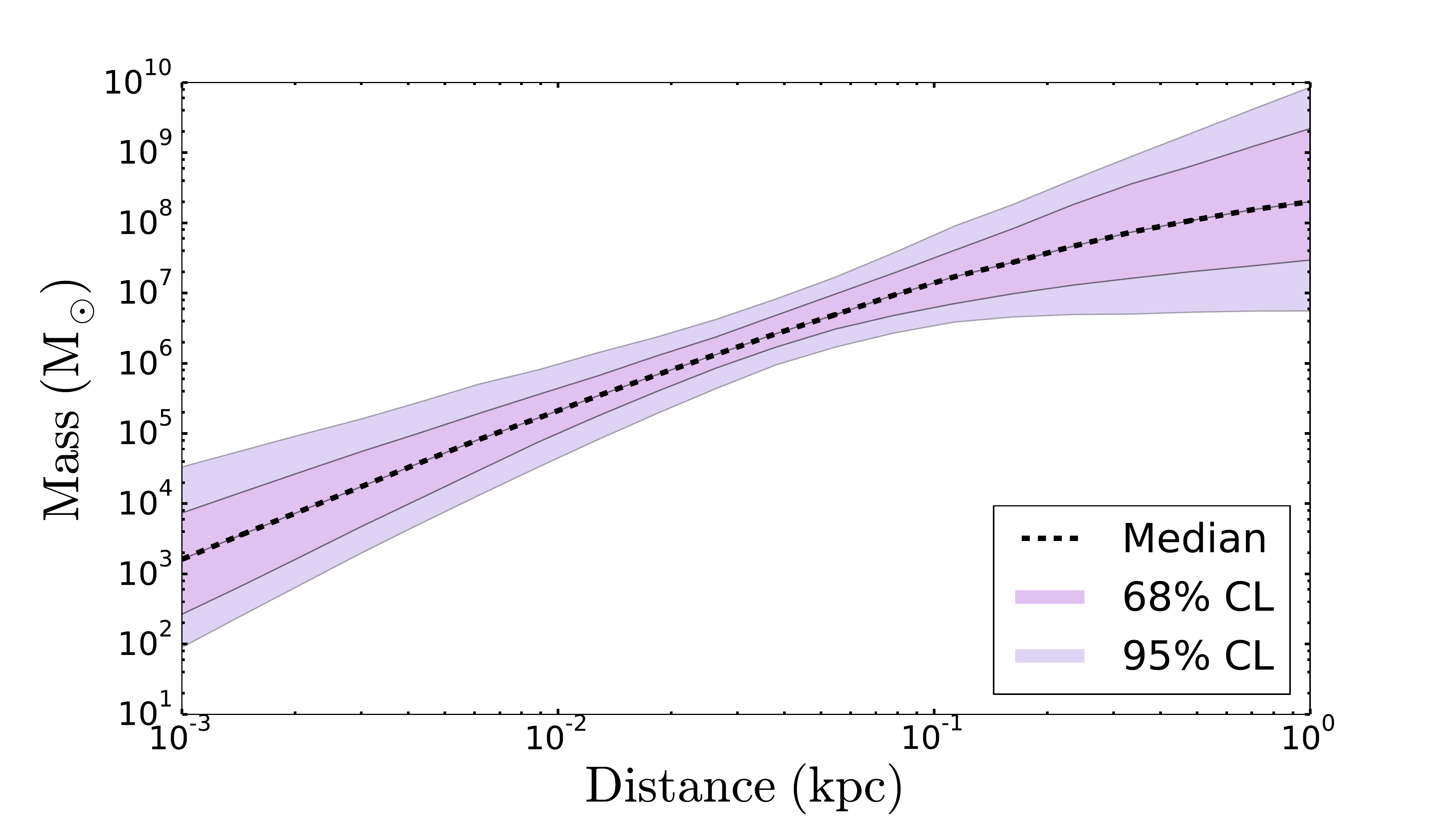}

		\caption{\it Mass as a function or radius for Triangulum II from the MCMC Jeans analysis outlined in the text.  The reader should note that the outermost star in the analysis is only 80.4pc from the centre of the halo, so the masses on the right of the diagram are simply extrapolations, see text.}

		\label{trimass}

	\end{figure}
The results of the Jeans analysis are plotted in Figure \ref{trimass}.  In this diagram, we integrated the best fitting density profiles out to 1 kpc and show what the mass of the halo would be at all radii up to that distance.  A halo such as Triangulum II, which is relatively close to the Milky Way and is also moving quickly, is likely to have suffered significant tidal stripping from its outer regions, so the mass estimates at large radii may well be an overestimate.  It is important to remember that the largest radius star included in the Jeans analysis is at 80.4 pc and that any mass estimate beyond that is simply an extrapolation of the density profile at low radii.  Nevertheless our analysis suggests that the enclosed mass at the radius of the outermost star is greater than $10^6 M_\odot$.  More precisely, the obtained half light mass is $2.15^{+1.42}_{-0.98} \times 10^6$ $M_{\odot}$ for the 68\% confidence interval and $2.15^{+4.56}_{-1.34} \times 10^6$ $M_{\odot}$ for the 95\% interval.

\section{Computing the J-factor and comparison to Milky Way satellites.}\label{jfac}
One of the many reasons why this halo may be of interest is because of its potential to help in the search for self-annihilating dark matter.  The flux of annihilating dark matter with a density distribution $\rho(r)$ is described by:
\begin{equation}
\frac{dF(\mathbf{\hat{n}}, E)}{d\Omega dE} = \frac{\langle \sigma v \rangle}{8 \pi M^{2}_{\chi}}\frac{dN_{\gamma}(E)}{dE} \int_{l=0}^{\infty}dl \left[\rho(l\mathbf{\hat{n}})\right]^{2}
\end{equation}
where the integral is performed over the line of sight. The integral is the J-factor or Astrophysical factor per solid angle $\Omega$ \citep{jfactor} which is essentially an integral of the density squared of dark matter within a cone along the line of sight with a particular opening angle - the relevant quantity for a substance annihilating with itself. The Particle Physics factor is dependent on the model of dark matter i.e its annihilation cross-section $\langle \sigma v \rangle$ and the mass of dark matter $M_{\chi}$.  A given particle physics model can be tested by obtaining these quantities from theory and combining them with the astrophysical J-factor, which is the subject of this study.

\begin{figure}

		  \includegraphics[trim=1.5cm 1.5cm 2.5cm 0.5cm,width=\columnwidth]{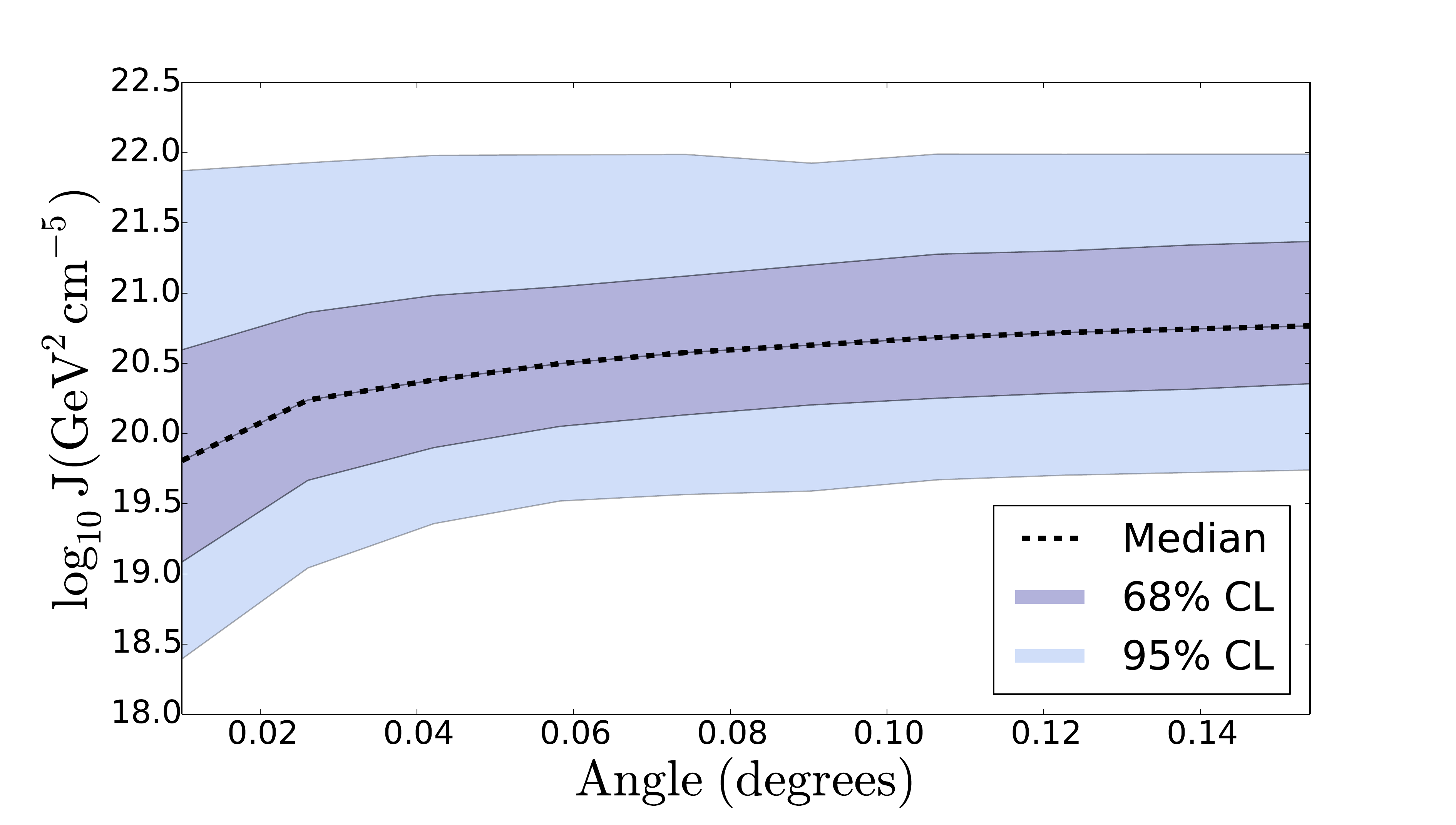}

		\caption{\it The J-factor for Triangulum II as a function of angle based upon the MCMC Jeans analysis decribed in the text.  We assume a density profile with $\alpha$, $\beta$ and $\gamma$ free parameters.  We only integrate out to the angle corresponding to the outermost star, 80.4 pc.  The line of sight integral also only includes the central 80.4 pc of the halo.  The two bands represent the 68\% and 95\% intervals and the central line is the median.}

		\label{trijfac}

	\end{figure}

One can obtain the total J-factor by integrating $ dJ( \theta ) / d \Omega$ over an angle $\theta_{max}$:
\begin{equation}
J(\theta_{max}) = \int_{0}^{\theta_{max}} \frac{dJ(\theta ')}{d\Omega}2\pi sin(\theta')d\theta'
\end{equation}
where $\theta_{max} = sin^{-1} (r_{max}/D)$ with D - the distance to the galaxy and $r_{max}$ - distance to furthest member star from the center of the galaxy \citep{jfactor}. We take this furthest star to be Star 14 in Table II and obtain $\theta_{max} = 0.15^{\circ}$.
We calculate the J-factors for the dark matter profiles produced by CLUMPY for angles up to $\theta_{max}$, using the values of $\rho_s$, $r_s$ produced from the CLUMPY MCMC. Figure \ref{trijfac} shows the median, 68\% and 95\% CL for each. Keeping $\alpha$ and $\gamma$ free allows for both cuspy and cored profiles and so a wider spread in the J-factors.
Note this result is compatible with another parallel analysis of the same object which appeared in the literature as we were writing up our results \citep{scoop}.  We obtain $\log_{10}(J/$GeV$^2$cm$^{-5})$ of $20.77^{+0.60}_{-0.41}$ at 68\% CL and $20.77^{+1.22}_{-1.03}$ at 95\% CL at the angle of the furthest star in our sample.  The value at 68\% CL is within the error given by \citet{scoop}, $\log_{10}$ ($J$/GeV$^{2}$ cm$^{-5}$) = 20.25$^{+1.28}_{-1.56}$ and \citet{hutten} 20.9$^{+1.4}_{-1.2}$. The values in \citet{scoop} are quoted at 0.5 degrees, however in that work the authors still restrict the J-factor to the angle of the furthest star if that angle is below 0.5 degrees while \citet{hutten} also use CLUMPY and their value is actually quoted at 0.5 degrees.  The J-factor we obtain at 0.5 degrees is 21.03$^{+0.83}_{-0.57}$ GeV$^2$ cm$^{-5}$ at 68\% CL, based on an extrapolation of the profiles out to larger radii.  We note that N-body simulations of dwarfs close to the centre of galaxies exhibit steepening of the outer profile due to tidal stripping \citep{tidal}.

This result suggests that the J-factor associated with Triangulum II is potentially more than $10^{20}$GeV$^2$cm$^{-5}$ which makes it one of the largest J-factors of any of the classical or ultrafaint dwarfs \citep{membership,scoop}.    A recent analysis of 21 dwarf spheroidals \citep{membership} predicted that the three brightest J factors would be  $\log_{10}(J_{0.5}/$GeV$^2$cm$^{-5})$ = 20.00$^{+0.7}_{-0.5}$ for Ursa Major II, $19.7^{+0.8}_{-0.7}$ for Coma and 19.6$^{+1.2}_{-0.6}$ for Willman I.  The J-factor is potentially greater than that of the recently discovered Reticulum-II dwarf galaxy which has $\log_{10}(J_{0.5}/$GeV$^2$cm$^{-5})$ = 19.6$^{+1.0}_{-0.7}$ \citep{retic}.

Of course, as we have repeatedly stressed, this analysis is based upon only a very small number of stars, and it is vitally important that a larger sample of stars are identified and their line of sight velocities measured to determine whether the apparently inflated velocity dispersion in the outer parts of the galaxy is the result of the effects of a largely dominant dark matter content, as it currently appears to be with so few stars. Looking for evidence of stellar streams could shed light onto whether this object has been tidally disrupted and its current orbit, or whether it is a stable dwarf galaxy. These further studies are essential before our presented J-factor estimates, complementary to those of \citet{scoop} and \citet{hutten}, can be used to inform observational strategies for upcoming searches such as those to be carried out by CTA.

\section{Conclusion and Discussion}

In this work we have looked at the dwarf galaxy Triangulum II which has recently been discovered in the Pan-STARRS survey.  There is very little data available for the stellar kinematics inside this dwarf and we have looked at 14 stars.

We first searched for signs of the dwarf galaxy being tidally disrupted by comparing the product of its diameter with the spatial velocity gradient to the overall velocity dispersion.  We showed using Monte Carlo realisations of fake data that, while the velocity gradient is quite large compared to the velocity dispersion, for such a small number of stars the relatively large value of the gradient is still statistically insignificant. 

We then proceeded to fit the underlying gravitational potential using a Jeans analysis to try to model the dark matter halo.  Having done that we obtained the J-factor, the relevant quantity for looking at the indirect detection of self-annihilating dark matter.  We found values for the J-factor between $10^{20}$ and $10^{22}$ GeV$^2$ cm$^{-5}$, which if they turned out to be reliable estimates, would make Triangulum II one of the best targets for searches for dark matter self annihilation in the sky.

This result is based upon very few stars and we very much hope that more line of sight velocities will be obtained for this object.  Tests could then be done on the velocity gradient with much smaller error bars, which could possibly show signs that the velocity dispersion of the object is due to tidal disruption rather than gravitational potential.  Alternatively, more stars could add to the hypothesis that the velocity dispersion is due to a dense dark matter core.  If this is the case, Triangulum II may turn out to be a very important object for astro-particle physics.

\section{Acknowledgments}
We are extremely grateful for conversations with Mark Birkinshaw, Vincent Bonnivard, Michelle Collins, Celine Combet, Michael Daniel, Jorge Penarubbia and Justin Read. MF acknowledges support from the STFC and the European Research Council under the European Union's Horizon 2020 program (ERC Grant Agreement no.648680) and is grateful for support from the IPPP in the form of an associateship.




\bibliographystyle{mnras}
\bibliography{triangulum} 

\begin{thebibliography}{}
\makeatletter
\relax
\def\mn@urlcharsother{\let\do\@makeother \do\$\do\&\do\#\do\^\do\_\do\%\do\~}
\def\mn@doi{\begingroup\mn@urlcharsother \@ifnextchar [ {\mn@doi@}
  {\mn@doi@[]}}
\def\mn@doi@[#1]#2{\def\@tempa{#1}\ifx\@tempa\@empty \href
  {http://dx.doi.org/#2} {doi:#2}\else \href {http://dx.doi.org/#2} {#1}\fi
  \endgroup}
\def\mn@eprint#1#2{\mn@eprint@#1:#2::\@nil}
\def\mn@eprint@arXiv#1{\href {http://arxiv.org/abs/#1} {{\tt arXiv:#1}}}
\def\mn@eprint@dblp#1{\href {http://dblp.uni-trier.de/rec/bibtex/#1.xml}
  {dblp:#1}}
\def\mn@eprint@#1:#2:#3:#4\@nil{\def\@tempa {#1}\def\@tempb {#2}\def\@tempc
  {#3}\ifx \@tempc \@empty \let \@tempc \@tempb \let \@tempb \@tempa \fi \ifx
  \@tempb \@empty \def\@tempb {arXiv}\fi \@ifundefined
  {mn@eprint@\@tempb}{\@tempb:\@tempc}{\expandafter \expandafter \csname
  mn@eprint@\@tempb\endcsname \expandafter{\@tempc}}}

\bibitem[\protect\citeauthoryear{{Abazajian} et~al.,}{{Abazajian}
  et~al.}{2009}]{SDSS}
{Abazajian} K.~N.,  et~al., 2009, \mn@doi [Astrophys. J.]
  {10.1088/0067-0049/182/2/543}, \href
  {http://adsabs.harvard.edu/abs/2009ApJS..182..543A} {182, 543}

\bibitem[\protect\citeauthoryear{Abbott et~al.}{Abbott et~al.}{2005}]{DES}
Abbott T.,  et~al., 2005

\bibitem[\protect\citeauthoryear{Ackermann et~al.}{Ackermann
  et~al.}{2015}]{fermi}
Ackermann M.,  et~al., 2015, \mn@doi [Phys. Rev. Lett.]
  {10.1103/PhysRevLett.115.231301}, 115, 231301

\bibitem[\protect\citeauthoryear{Actis et~al.}{Actis et~al.}{2011}]{CTA}
Actis M.,  et~al., 2011, \mn@doi [Exper. Astron.] {10.1007/s10686-011-9247-0},
  32, 193

\bibitem[\protect\citeauthoryear{Bechtol et~al.}{Bechtol
  et~al.}{2015}]{bechtol}
Bechtol K.,  et~al., 2015, \mn@doi [Astrophys. J.]
  {10.1088/0004-637X/807/1/50}, 807, 50

\bibitem[\protect\citeauthoryear{Belokurov et~al.}{Belokurov
  et~al.}{2006}]{bootes}
Belokurov V.,  et~al., 2006, \mn@doi [Astrophys. J.] {10.1086/507324}, 647,
  L111

\bibitem[\protect\citeauthoryear{Bergstr{\"o}m, Ullio  \&
  Buckley}{Bergstr{\"o}m et~al.}{1998}]{lars}
Bergstr{\"o}m L.,  Ullio P.,   Buckley J.~H.,  1998, \mn@doi [Astropart. Phys.]
  {10.1016/S0927-6505(98)00015-2}, 9, 137

\bibitem[\protect\citeauthoryear{{Binney} \& {Merrifield}}{{Binney} \&
  {Merrifield}}{1998}]{binneymerrifield}
{Binney} J.,  {Merrifield} M.,  1998, {Galactic Astronomy}

\bibitem[\protect\citeauthoryear{Bonnivard, Combet, Maurin  \&
  Walker}{Bonnivard et~al.}{2015a}]{bonni15}
Bonnivard V.,  Combet C.,  Maurin D.,   Walker M.~G.,  2015a, \mn@doi [Mon.
  Not. Roy. Astron. Soc.] {10.1093/mnras/stu2296}, 446, 3002

\bibitem[\protect\citeauthoryear{Bonnivard, Combet, Maurin  \&
  Walker}{Bonnivard et~al.}{2015b}]{Input}
Bonnivard V.,  Combet C.,  Maurin D.,   Walker M.~G.,  2015b, \mn@doi [Mon.
  Not. Roy. Astron. Soc.] {10.1093/mnras/stu2296}, 446, 3002

\bibitem[\protect\citeauthoryear{Bonnivard et~al.}{Bonnivard
  et~al.}{2015c}]{membership}
Bonnivard V.,  et~al., 2015c, \mn@doi [Mon. Not. Roy. Astron. Soc.]
  {10.1093/mnras/stv1601}, 453, 849

\bibitem[\protect\citeauthoryear{Bonnivard et~al.,}{Bonnivard
  et~al.}{2015d}]{retic}
Bonnivard V.,  et~al., 2015d, \mn@doi [Astrophys. J.]
  {10.1088/2041-8205/80/2/L36}, 808, L36

\bibitem[\protect\citeauthoryear{Bonnivard, H{\"u}tten, Nezri, Charbonnier,
  Combet  \& Maurin}{Bonnivard et~al.}{2016}]{CLUMPY}
Bonnivard V.,  H{\"u}tten M.,  Nezri E.,  Charbonnier A.,  Combet C.,   Maurin
  D.,  2016, \mn@doi [Comput. Phys. Commun.] {10.1016/j.cpc.2015.11.012}, 200,
  336

\bibitem[\protect\citeauthoryear{Drlica-Wagner et~al.}{Drlica-Wagner
  et~al.}{2015}]{drilca}
Drlica-Wagner A.,  et~al., 2015, \mn@doi [Astrophys. J.]
  {10.1088/0004-637X/813/2/109}, 813, 109

\bibitem[\protect\citeauthoryear{Evans, An  \& Walker}{Evans
  et~al.}{2009}]{los}
Evans N.~W.,  An J.,   Walker M.~G.,  2009, \mn@doi [Mon. Not. Roy. Astron.
  Soc.] {10.1111/j.1745-3933.2008.00596.x}, 393, 50

\bibitem[\protect\citeauthoryear{Geringer-Sameth, Koushiappas  \&
  Walker}{Geringer-Sameth et~al.}{2015}]{jfactor}
Geringer-Sameth A.,  Koushiappas S.~M.,   Walker M.,  2015, \mn@doi [Astrophys.
  J.] {10.1088/0004-637X/801/2/74}, 801, 74

\bibitem[\protect\citeauthoryear{Hayashi, Ichikawa, Matsumoto, Ibe, Ishigaki
  \& Sugai}{Hayashi et~al.}{2016}]{scoop}
Hayashi K.,  Ichikawa K.,  Matsumoto S.,  Ibe M.,  Ishigaki M.~N.,   Sugai H.,
  2016

\bibitem[\protect\citeauthoryear{Hooper \& Linden}{Hooper \&
  Linden}{2015}]{linden}
Hooper D.,  Linden T.,  2015, \mn@doi [JCAP] {10.1088/1475-7516/2015/09/016},
  1509, 016

\bibitem[\protect\citeauthoryear{H{\"u}tten, Combet, Maier  \&
  Maurin}{H{\"u}tten et~al.}{2016}]{hutten}
H{\"u}tten M.,  Combet C.,  Maier G.,   Maurin D.,  2016

\bibitem[\protect\citeauthoryear{Kaiser et~al.}{Kaiser et~al.}{2002}]{kaiser}
Kaiser N.,  et~al., 2002, \mn@doi [Proc. SPIE Int. Soc. Opt. Eng.]
  {10.1117/12.457365}, 4836, 154

\bibitem[\protect\citeauthoryear{{Kent}}{{Kent}}{1990}]{Kent}
{Kent} S.~M.,  1990, Mon. Not. Roy. Astron. Soc., \href
  {http://adsabs.harvard.edu/abs/1990MNRAS.247..702K} {247, 702}

\bibitem[\protect\citeauthoryear{{Kim} \& {Jerjen}}{{Kim} \&
  {Jerjen}}{2015}]{kimjerjen}
{Kim} D.,  {Jerjen} H.,  2015, \mn@doi [Astrophys. J.]
  {10.1088/2041-8205/808/2/L39}, \href
  {http://adsabs.harvard.edu/abs/2015ApJ...808L..39K} {808, L39}

\bibitem[\protect\citeauthoryear{Kirby, Cohen, Simon  \& Guhathakurta}{Kirby
  et~al.}{2015}]{kirby}
Kirby E.~N.,  Cohen J.~G.,  Simon J.~D.,   Guhathakurta P.,  2015, \mn@doi
  [Astrophys. J.] {10.1088/2041-8205/814/1/L7}, 814, L7

\bibitem[\protect\citeauthoryear{Klessen \& Kroupa}{Klessen \&
  Kroupa}{1998}]{falsealarm}
Klessen R.,  Kroupa P.,  1998, \mn@doi [Astrophys. J.] {10.1086/305540}, 498,
  143

\bibitem[\protect\citeauthoryear{Klypin, Kravtsov, Valenzuela  \& Prada}{Klypin
  et~al.}{1999}]{missingsats}
Klypin A.~A.,  Kravtsov A.~V.,  Valenzuela O.,   Prada F.,  1999, \mn@doi
  [Astrophys. J.] {10.1086/307643}, 522, 82

\bibitem[\protect\citeauthoryear{{Laevens} et~al.,}{{Laevens}
  et~al.}{2015}]{Leavens}
{Laevens} B.~P.~M.,  et~al., 2015, \mn@doi [Astrophys. J.]
  {10.1088/2041-8205/802/2/L18}, \href
  {http://adsabs.harvard.edu/abs/2015ApJ...802L..18L} {802, L18}

\bibitem[\protect\citeauthoryear{{Martin} et~al.,}{{Martin}
  et~al.}{2016}]{hotstellar}
{Martin} N.~F.,  et~al., 2016, \mn@doi [Astrophys. J.]
  {10.3847/0004-637X/818/1/40}, \href
  {http://adsabs.harvard.edu/abs/2016ApJ...818...40M} {818, 40}

\bibitem[\protect\citeauthoryear{Moore, Ghigna, Governato, Lake, Quinn, Stadel
  \& Tozzi}{Moore et~al.}{1999}]{lcdm}
Moore B.,  Ghigna S.,  Governato F.,  Lake G.,  Quinn T.~R.,  Stadel J.,
  Tozzi P.,  1999, \mn@doi [Astrophys. J.] {10.1086/312287}, 524, L19

\bibitem[\protect\citeauthoryear{{Niederste-Ostholt}, {Belokurov}, {Evans},
  {Gilmore}, {Wyse}  \& {Norris}}{{Niederste-Ostholt} et~al.}{2009}]{segtid}
{Niederste-Ostholt} M.,  {Belokurov} V.,  {Evans} N.~W.,  {Gilmore} G.,  {Wyse}
  R.~F.~G.,   {Norris} J.~E.,  2009, \mn@doi [\mnras]
  {10.1111/j.1365-2966.2009.15287.x}, \href
  {http://adsabs.harvard.edu/abs/2009MNRAS.398.1771N} {398, 1771}

\bibitem[\protect\citeauthoryear{Pe{\~n}arrubia, Navarro, McConnachie  \&
  Martin}{Pe{\~n}arrubia et~al.}{2009}]{tidal}
Pe{\~n}arrubia J.,  Navarro J.~F.,  McConnachie A.~W.,   Martin N.~F.,  2009,
  \mn@doi [Astrophys. J.] {10.1088/0004-637X/698/1/222}, 698, 222

\bibitem[\protect\citeauthoryear{{Pryor}}{{Pryor}}{1996}]{pryor}
{Pryor} C.,  1996, in {Morrison} H.~L.,  {Sarajedini} A.,  eds,  Astronomical
  Society of the Pacific Conference Series Vol. 92, Formation of the Galactic
  Halo...Inside and Out. p.~424

\bibitem[\protect\citeauthoryear{Putze \& Derome}{Putze \&
  Derome}{2014}]{GreAT}
Putze A.,  Derome L.,  2014, \mn@doi [Phys. Dark Univ.]
  {10.1016/j.dark.2014.07.002}, 5-6, 29

\bibitem[\protect\citeauthoryear{Read, Agertz  \& Collins}{Read
  et~al.}{2016}]{baryonicfeedback}
Read J.~I.,  Agertz O.,   Collins M. L.~M.,  2016, \mn@doi [Mon. Not. Roy.
  Astron. Soc.] {10.1093/mnras/stw713}, 459, 2573

\bibitem[\protect\citeauthoryear{Richardson \& Fairbairn}{Richardson \&
  Fairbairn}{2014}]{Richardson}
Richardson T.,  Fairbairn M.,  2014, \mn@doi [Mon. Not. Roy. Astron. Soc.]
  {10.1093/mnras/stu691}, 441, 1584

\bibitem[\protect\citeauthoryear{Schneider}{Schneider}{2014}]{schneider}
Schneider P.,  2014, Extragalactic astronomy and cosmology: an introduction.
Springer

\bibitem[\protect\citeauthoryear{{Walker} \& {Pe{\~n}arrubia}}{{Walker} \&
  {Pe{\~n}arrubia}}{2011}]{gaia}
{Walker} M.~G.,  {Pe{\~n}arrubia} J.,  2011, \mn@doi [\apj]
  {10.1088/0004-637X/742/1/20}, \href
  {http://adsabs.harvard.edu/abs/2011ApJ...742...20W} {742, 20}

\bibitem[\protect\citeauthoryear{Walker, Mateo, Olszewski, Bernstein, Wang  \&
  Woodroofe}{Walker et~al.}{2006}]{walkererrors}
Walker M.~G.,  Mateo M.,  Olszewski E.~W.,  Bernstein R.~A.,  Wang X.,
  Woodroofe M.,  2006, \mn@doi [Astron. J.] {10.1086/500193, 10.1086/505303},
  131, 2114

\bibitem[\protect\citeauthoryear{Wolf, Martinez, Bullock, Kaplinghat, Geha,
  Munoz, Simon  \& Avedo}{Wolf et~al.}{2010}]{Wolf}
Wolf J.,  Martinez G.~D.,  Bullock J.~S.,  Kaplinghat M.,  Geha M.,  Munoz
  R.~R.,  Simon J.~D.,   Avedo F.~F.,  2010, \mn@doi [Mon. Not. Roy. Astron.
  Soc.] {10.1111/j.1365-2966.2010.16753.x}, 406, 1220

\bibitem[\protect\citeauthoryear{Zhao}{Zhao}{1996}]{zhao}
Zhao H.,  1996, \mn@doi [Mon. Not. Roy. Astron. Soc.]
  {10.1093/mnras/278.2.488}, 278, 488

\bibitem[\protect\citeauthoryear{Wolf, Martinez, Bullock, Kaplinghat, Geha,
  Munoz, Simon  \& Avedo}{gai}{}]{gaiawiki}
Gaia Challenge Wiki,
  \url{http://astrowiki.ph.surrey.ac.uk/dokuwiki/doku.php?id=tests:sphtri}

\makeatother
\end{thebibliography}

\bsp	
\label{lastpage}
\end{document}